\documentstyle[preprint,aps]{revtex}
\newtheorem{teo}{Theorem}
\newtheorem{prop}[teo]{Proposition}
\newtheorem{cor}[teo]{Corollary}
\def\K{{\cal K}}
\def\Rbar{\overline{R}}

\def\ptilde{\widetilde{p}}
\def\Gr{\widetilde{\Gamma}}
\def\D{{\cal D}}
\def\A{{\cal A}}
\def\S{\Sigma}
\def\Ds{{\cal D}\!\!\!/\:}
\def\ns{n^*}

\def\d{\partial}
\def\Tr{{\rm Tr}}
\def\L{{\cal L}}

\begin{document}
\draft
\title{Renormalization of gauge invariant composite operators in light-cone
       gauge}
\author{C. Acerbi and  A. Bassetto}
\address{Dipartimento di Fisica, Universit\`a di Padova and \\
	   INFN, Sezione di Padova, 35100 Padova, Italy}
\maketitle
\begin{abstract}
We generalize to composite operators concepts and techniques which
have been successful in proving renormalization of the effective
Action in light-cone gauge. Gauge invariant operators can be grouped
into classes, closed under renormalization, which is matrix-wise.
In spite of the presence of non-local counterterms,
an ``effective" dimensional
hierarchy still guarantees that any class is endowed with a finite
number of elements. The main result we find is that gauge
invariant operators under renormalization mix only among themselves,
thanks to the
very simple structure of Lee-Ward identities in this gauge, contrary
to their behaviour in covariant gauges.

\end{abstract}

\vspace{2cm}

\hfill {\large DFPD 93/TH/53}

\hfill {July 1993}

\vfill \eject

\section{INTRODUCTION}

Composite operators often occur in calculations of physical
cross-sections. A celebrated example is deep-inelastic scattering
where short-distance products of currents are expressed in terms
of local operators by means of a Wilson expansion \cite{1}. But, strictly
speaking, the lagrangian density itself is an instance of composite
operator.

As quantum fields are distribution-valued operators, one can easily
realize that, taking products at the same space-time point, gives rise
to singularities. Whence the need of first considering a procedure
of regularization and then performing the necessary subtractions in
a consistent way to operatively define their finite parts \cite{2}, at
least in a perturbative context.

The peculiar phenomenon occurring in composite operator renormalization
is their {\it mixing}.
Locality and polynomiality in the masses of counterterms guarantee
the presence of a {\it
dimensional hierarchy}: counterterms can only have canonical
dimensions less than or equal to the ones of the operators
we are considering. Therefore,
the number of counterterms which mix is finite \cite{2}.

All those concepts and techniques naturally apply to gauge theories
with the proviso they have to comply with Ward identities taking
care of redundant degrees of freedom. In covariant gauges (typically
in generalized Feynman gauges) the relevant Slavnov-Taylor identities
involve unphysical operators (Faddeev-Popov ghosts). As a consequence,
a deep thorough analysis \cite{3} has shown that gauge invariant operators
{\it do mix} with unphysical ones under renormalization.

The situation radically changes in the so-called {\it physical}
gauges $n_\mu A^\mu =0$, $n_\mu$ being a constant vector, where
there is no need of Faddeev-Popov fields and Lee-Ward identities
are straightforward \cite{4}. This is the reason why such gauges have been
largely adopted in the past for phenomenological applications \cite{5}.

Only recently however a systematic approach has been developed
with a sound basis on the axioms of canonical quantum field theory.
Effective Action renormalization has been proven, so far, at any
order in the loop expansion,  only in light-cone (LC) gauge
($n^2=0$)  \cite{6}.
Essential to this goal is to endow
the ``spurious" singularity, occurring in the vector propagator,
with a causal prescription (Mandelstam-Leibbrandt (ML) prescription
 \cite{7,8}), as suggested by a careful canonical quantization  \cite{9}.
This prescription in turn is the source of a potentially serious
difficulty: non-local counterterms are needed, already at one loop
level, to make one particle irreducible vertices finite  \cite{8}.

It is clear that non-locality could in principle destroy dimensional
hierarchy. Should the mixing involve an infinite number of
independent counterterms, even for a single insertion, the very
program of composite operator renormalization would be in jeopardy.

Happily this is not the case. Generalizing concepts and techniques
which have been successful in proving renormalization of the effective
Action,
in next Sections we show that a new kind of ``effective" dimensional hierarchy
can be established which is enough to prove renormalization at any
order in the loop expansion, at
least for gauge invariant composite operators, which are the ones
directly involved in phenomenological applications  \cite{10}. Actually the
very
simple structure of Lee-Ward identities, which survives renormalization
in this case, will allow us to reach a rather strong result:
in LC gauge, under renormalization gauge invariant operators mix
only among themselves, in classes with finite numbers of elements.

The problems one encounters when treating
more general operators, will be briefly discussed in the Conclusions.

In Sect. \ref{2}  we introduce our notation, we define the generating
functionals with composite operators insertions and derive the
Lee-Ward identities they have to satisfy. Sect. \ref{3} is devoted to
generalize BPHZ subtraction method \cite{11,12,13}
to our problem and prove the gauge invariance of renormalized composite
operators. In Sect. \ref{4}
we discuss power-counting in LC gauge and the need of introducing
a more general criterion of superficial degree of divergence, in
relation to Weinberg's theorem  \cite{14}. In Sect. \ref{5} we explore all
constraints the counterterms have to fulfill and  in Sect.\ref{6} we
prove that in
the mixing of gauge invariant operators a unique independent
non-local structure can appear with mass dimension equal to $one$,
the same one encounters when renormalizing the effective Action.
Concrete examples of
mixing are presented in Sect. \ref{7}, while remarks and comments concerning
further developments are contained in the Conclusions.

\section{THE GENERATING FUNCTIONALS WITH COMPOSITE OPERATORS} \label{2}
We start by defining our lagrangian and our notation
\begin{equation}
   \L_{g.i.}=- {1\over 2} \Tr \left( F_{\mu\nu}F^{\mu\nu} \right)
   + \bar\psi \left(i \Ds -m \right) \psi    \label{Lgi}
\end{equation}
where $F_{\mu\nu}$ is the usual field tensor in the adjoint representation
of the algebra su(N).
\begin{eqnarray}
&&    F_{\mu\nu}  =  \d_\mu A_\nu - \d_\nu A_\mu - ig [A_\mu,A_\nu] \: ,  \\
&&    A_\mu = A_\mu^a \tau^a, \:\:\:\:\: a=1,\ldots,N^2-1  \: ,  \\
&&    \Tr(\tau^a\tau^b) = {1\over 2}  \delta^{ab}\: , \\
&&    \left[ \tau^a , \tau^b \right] = i f^{abc} \tau^c \: ,
\end{eqnarray}
$f^{acb}$
being the structure constants of the group which are completely
antisymmetric in this basis. $\D_\mu$ is the covariant derivative acting on
the fundamental representation,
\begin{equation}
   \D_\mu = \d_\mu - i g A_\mu \: .
\end{equation}
The lagrangian density in eq. (\ref{Lgi}) is invariant under gauge
transformations, as is well known. Their infinitesimal form is
 \begin{mathletters}
\begin{eqnarray}
&& \delta^{[\omega]} \psi (x) = i g\, \omega(x)  \psi (x)  \: ,\\
&&  \delta^{[\omega]} A_\mu(x) = D_\mu \omega(x)  \:,
\end{eqnarray}
\end{mathletters}
where $D_\mu$  is the covariant derivative acting on the adjoint
representation
\begin{equation}
D_\mu = \d_\mu - ig[A_\mu,\:\cdot\:]
\end{equation}
and $\omega(x)$ are the infinitesimal parameters of the transformation.
In order to quantize the theory, we introduce the light-cone gauge fixing
\begin{equation}
 \L_{g.f.} = - \lambda(x) \cdot n_\mu A^\mu(x) \: ,   \label{Lgf}
\end{equation}
$\lambda(x)$ being Lagrange multipliers and $n_\mu$ a fixed light-like
four-vector $n^2=0$.

In the following, dimensional regularization will be understood in
the framework of perturbation theory. In $2\omega$ dimensions
the coupling constant
$g$ will be replaced with $g\mu^{2-\omega}$ where $\mu$ is a mass scale.

{}From eqs. (\ref{Lgi}) and (\ref{Lgf})  we can construct the usual functional
$W$ which generates the Green's functions of the theory
\begin{eqnarray}
W[J,K,\eta,\bar\eta] = {\cal N} \int [dA][d\lambda][d\psi][d\bar\psi]
exp \left[i \int \!\!d^{4}x \left(\L_{g.i.} + \L_{g.f.} + \L_s
 \right)\right]
\end{eqnarray}
where
\begin{equation}
   \L_s = J_\mu \cdot A^\mu + K \cdot \lambda + \bar\eta \psi -
   \bar\psi\eta.
\end{equation}
Then we can define in the usual way the functional $Z= {1\over i} \log W$
which generates the connected Green's functions; from $Z$ we get  the
``classical '' fields  $\A_\mu$, $\Lambda$, $\Psi$, $\bar\Psi$ and,
eventually, the functional $\Gamma$ which generates the
proper vertices of the theory
\begin{eqnarray}
\Gamma[\A,\Lambda,\Psi,\bar\Psi] =
Z[J,K,\eta,\bar\eta]
- \int d^{4}x
\left( J \cdot \A + K\cdot \Lambda + \bar\Psi \eta - \bar\eta\Psi \right).
\end{eqnarray}
The derivatives with respect to Grassmann variables are understood as left
derivatives; with our conventions we get, in particular,
\begin{equation}
\left\{
  \begin{array}{rcl}
     \bar\Psi \equiv \displaystyle\frac{\delta Z}{\delta \eta}
\rule[-5mm]{0mm}{5mm}, \\
     \Psi \equiv \displaystyle\frac{\delta Z}{\delta \bar\eta}
\rule{0mm}{5mm},
  \end{array} \right.
\hspace{1cm}  \Rightarrow
\hspace{1cm}
\left\{
  \begin{array}{rcl}
     \bar\eta = \displaystyle\frac{\delta \Gamma}{\delta \Psi}
\rule[-5mm]{0mm}{5mm}, \\
     \eta = \displaystyle\frac{\delta \Gamma}{\delta \bar\Psi}
\rule{0mm}{5mm}.
  \end{array} \right.
\end{equation}

We also notice that invariance under a shift in $\lambda$ of the path-
integral entails the condition
\begin{equation}  \label{14}
n\A=K ,
\end{equation}
which in turn guarantees that
any Green's function containing $nA$, but no $\lambda$, vanishes.

In this section we are mainly concerned with the generalization of such
generating functionals to the case
in which composite operators are considered.
Such a generalization is presented for instance in refs. \cite{15,16}.
We denote by $X=X[A,\psi,\bar\psi]$ a polynomial built
from the original fields and their
derivatives taken at the same space-time point; Green's functions
with insertions of such operators usually exhibit further singularities.
The technique one uses to take those insertions into account is to introduce,
in the definition of $W$, a source term related to $X$.
In the general case we shall consider
a set of operators $X_i$ each associated to a source $\S_i$, by adding
the lagrangian density
\begin{equation}
  \L_X = \sum_i \S_i \cdot X_i \: .
\end{equation}
In the following we shall not
be concerned with composite operators
involving Lagrange multipliers as they
would affect the equations of motion
of the field $\lambda$ that enter in the derivation of the Lee identities.
Moreover we shall limit ourselves to gauge invariant composite operators,
but in the final section where we shall briefly dwell on possible
generalizations.

A crucial point to remark is that the functional $\Gamma$ with insertions
is defined by means of a Legendre transformation involving only the classical
fields we have already considered
\begin{eqnarray}
\Gamma[\A,\Lambda,\Psi,\bar\Psi,\S_i] =
Z[J,K,\eta,\bar\eta,\S_i] - \int \!\!d^{4}x
\left( J \!\cdot \!\A + K\!\cdot\!\Lambda +
\bar\Psi \eta - \bar\eta\Psi \right).
\end{eqnarray}
As a consequence, one can prove the equality
\begin{equation}
\left.\frac{\delta\Gamma}{\delta\S_i}\right|_{\A,\Lambda,\Psi,\bar\Psi,
\S_{j\neq i}} =
\left.\frac{\delta Z}{\delta\S_i}\right|_{J,K,\eta,\bar\eta,\S_{j\neq i}}
  \: ,
\end{equation}
where in the left-hand side (right-hand side) ``classical'' fields
(original sources) are kept fixed besides the sources $\S_{j\neq i}$.

By solving the equations of motion of the Lagrange multiplier it is possible
to make explicit the dependence of $\Gamma$ on $\Lambda$
\begin{equation}
\Gamma[\A,\Lambda,\Psi,\bar\Psi,\S_i] =
\Gr[\A,\Psi,\bar\Psi,\S_i] -\int d^{4}x \: \Lambda\cdot n\A \:
\end{equation}
and to convince oneself that the gauge-fixing term does not renormalize.
$\Gr$ is customarily called the ``reduced generating functional''.

As we are concerned with gauge-invariant operators, Lee-Ward identities, which
have a very simple form in  light-cone gauge, will not entail further
difficulties in presence of insertions.
In order to derive the Ward identities, we follow the standard technique of
performing a change of variable in the path integral corresponding to
an infinitesimal gauge transformation. The related functional determinant
in this gauge is independent of the fields, as is well known.
As we are here considering
gauge-invariant insertions, they cannot affect the form of the Ward identity
\begin{eqnarray}
D_\mu^{ab}\left[ \frac{\delta}{i\delta J}\right]
 \left\{
\frac{\delta}{i\delta K^b} n^\mu - J^{\mu b}
\right\} W \label{Ward} +
ig\mu^{2-\omega}
\left(
\bar\eta \tau^a \frac{\delta W}{i\delta\bar\eta} +
\frac{\delta W}{i\delta\eta}  \tau^a \eta
\right) =0 ,
\end{eqnarray}
where $W$ depends also on sources $\S_i$ related to composite operators.
We can get rid of the term with second order functional derivative in eq.
(\ref{Ward}), using the equations of motion for the Lagrange multiplier.
Then we derive from eq. (\ref{Ward}) the following Lee identity for
the reduced functional $\Gr$
\begin{equation}
D_\mu^{ab}\left[ \A \right] \frac{\delta\Gr}{\delta \A_\mu^b}
+ ig\mu^{2-\omega} \left( \frac{\delta\Gr}{\delta \Psi}  \tau^a \Psi
+ \bar\Psi \tau^a
\frac{\delta\Gr}{\delta \bar\Psi}
\right)
\equiv \Delta^a\Gr =0 ,   \label{Lee}
\end{equation}
$\Delta^a$ being the functional differential operator which describes
an infinitesimal gauge transformation of the ``classical'' fields.
We shall use the same symbol also for the analogous operator
acting on functionals of elementary fields.
Eq. (\ref{Lee})  means that $\Gr$ is gauge invariant.
We stress the fact that $\Gr$ depends also on possible sources
related to gauge-invariant composite operators.

\section{GAUGE INVARIANCE OF RENORMALIZED OPERATORS} \label{3}
In order to renormalize either the Action or a composite operator, we
adopt the graph-by-graph subtraction method (or BPHZ method) summarized by the
Bogoliubov's $\Rbar$ operator on Feynman graphs \cite{11,12,13}.
We just stress the fact here that in presence of diagrams with operator
insertions the definition of 1PI diagram remains the same if the operator
vertices are treated just like ordinary interaction vertices.

In the following we shall work in the minimal subtraction scheme  (MS)
on dimensionally regularized diagrams: we denote with $\K G$ the singular
part of the Laurent expansion of the graph $G$ in a neighborhood of
$\omega = 2$.
The renormalized graph $RG$ is obtained by   subtracting the singular part
from the subdivergence-free diagram $\Rbar G$
\begin{equation}
	RG= (1-\K) \Rbar G \: .
\end{equation}
We shall also use the notation
\begin{equation}
	CG= -\K \Rbar G
\end{equation}
to indicate the specific counterterm necessary to renormalize the graph $G$.
$CG$ is different from zero if and only if $G$ is 1PI and superficially
divergent. In this case a specific counterterm chosen to produce $CG$ as an
additional Feynman rule, has to be added to the lagrangian. If $G$
involves composite operator vertices
it contributes to the renormalization of the
$\S$-dependent terms, otherwise it renormalizes the original lagrangian.

By performing this procedure for every 1PI diagram up to order l-loops, one
builds an Action $S^{\{l\}}$ renormalized  to this order.
A synthetic and completely equivalent description of this method is given
through the generating functional $\Gamma$.
We define
\begin{eqnarray}           \label{S0}
S^{\{0\}}[A,\psi,&\bar\psi&,\S_i] =
\int \!\! d^{4}x \left( \L_{g.i.}[A,\psi,\bar\psi]
+ \L_X [A,\psi,\bar\psi,\S_i] \right),
\end{eqnarray}
as unrenormalized Action. In this definition the gauge-fixing term is
excluded as it does not renormalize; hence $S^{\{0\}}$ is gauge invariant.
We denote by $\Gr^{\{l\}}$ the reduced generating functional obtained from
the Action $S^{\{l\}}$ and perform the loopwise expansion
\begin{equation}
	 \Gr^{\{l\}} = \sum_{m=0}^\infty \Gr^{\{l\}}_m \: ;
\end{equation}
$\Gr^{\{l\}}_m$ represents the m-loops contribute to $\Gr^{\{l\}}$.
Now we are able to define iteratively the renormalized Action
\begin{eqnarray}
  S^{\{l\}}[A,\psi,\bar\psi,\S_i] =
  S^{\{l-1\}}[A,\psi,\bar\psi,\S_i] - \K
  \Gr^{\{l-1\}}_l [A,\psi,\bar\psi,\S_i]  \: ,     \label{Sl}
\end{eqnarray}
where $\K$ picks up just the singular part at $\omega=2$ of the regularized
expression $\Gr^{\{l-1\}}_l [A,\psi,\bar\psi,\S_i]$; in
this functional, the fields $A$, $\psi$ and $\bar\psi$
take the place of the corresponding classical fields.

In general, even in a covariant theory,
if the dimension of  $X_i$ is $\geq 4$, an infinite number of
counterterms of arbitrarily high degree in $\S_i$ are introduced in the
renormalized Action by eq. (\ref{Sl}).
The ``renormalized operator'' $[X_k]^{\{l\}}$ is defined by
\begin{equation}
  [X_k]^{\{l\}}(x) \equiv  \left.
 \frac{ \delta S^{\{l\}}[A,\psi,\bar\psi,\S_i]}{\delta \S_k(x)}
 \right|_{\S_i=0\:\forall i } \: ;
 \label{Xl}
\end{equation}
the operator is renormalized only in the sense that Green's functions
with at most one single insertion of $[X_k]^{\{l\}}$
are finite in the renormalized
theory. If finite Green's functions with more operator insertions are
needed one has to consider the whole renormalized Action
$S^{\{l\}}$ whose functional
$W^{\{l\}}[J,K,\eta,\bar\eta,\Sigma_i]$ is finite up
to order $l$ at any degree in $\S_i$.

A ``weak'' form of renormalization will also be considered in which only
counterterms at most linear in the sources $\S_i$ are introduced.
To this purpose we define
\begin{equation}
\Gr^L\equiv \left.\Gr\right|_{\S_i=0}
+ \int d^{4}x \: \sum_j \left( \left. \frac{\delta \Gr}{\delta \S_j(x)}
\right|_{\S_i=0} \cdot
\S_j(x) \right)
\end{equation}
the part of $\Gr$ linear in the sources $\S_i$.
Then we define  recursively as  weakly renormalized Action
\begin{eqnarray}
  S^{\{l\}}_w[A,\psi,\bar\psi,\S_i] =
  S^{\{l-1\}}_w[A,\psi,\bar\psi,\S_i] - \K
  \Gr^{L\{l-1\}}_l [A,\psi,\bar\psi,\S_i]  \: .     \label{Swl}
\end{eqnarray}
Of course one gets
\begin{equation}       \label{ofcourse}
  S^{\{l\}}_w[A,\psi,\bar\psi,\S_i] =
  S^{\{l\}}[A,\psi,\bar\psi] + \sum_i \S_i \cdot [X_i]^{\{l\}} \: ,
\end{equation}
where the first term in the r.h.s. is the renormalized Action one would
obtain if operator insertions were absent.
Only the linear parts in the $\S_i$s of the generating functionals
obtained from $S_w$ are finite.

We shall now prove the
\begin{prop} Let  $S^{\{0\}}$  be
the Action with insertions of gauge invariant operators
$X_i$, defined by eq. (\ref{S0}) and  $S^{\{l\}}$
the Action renormalized up to l-loops according to
eq. (\ref{Sl}), then
\begin{enumerate}
\item  $S^{\{l\}}$ is gauge invariant $\forall l$
\begin{equation}                          \label{DSl}
       \Delta^a S^{\{l\}}[A,\psi,\bar\psi] = 0.
\end{equation}
\item  the renormalized operators $[X_i]^{\{l\}}$ are gauge invariant
$\forall l$
\begin{equation}
       \Delta^a [X_i]^{\{l\}}[A,\psi,\bar\psi] = 0.
\end{equation}
\end{enumerate}
\end{prop}
{\bf proof:} {\em 2.} follows directly from {\em 1.} and eq. (\ref{Xl}).
Let us  show {\em 1.}: by induction on $l$; as it obviously holds
if $l=0$,
we have just to prove the inductive step.
Let us start from eq. (\ref{DSl}).
The form of Lee identities is not affected by renormalization
\begin{equation}
       \Delta^a \Gr^{\{l\}}[\A,\Psi,\bar\Psi] = 0.
\end{equation}
The same equation must hold for the singular part of the Laurent expansion
in $\omega =2$ and for each contribution in the loopwise expansion
\begin{equation}
       \Delta^a \K \Gr_m^{\{l\}}[\A,\Psi,\bar\Psi] = 0.
\end{equation}
{}From eq. (\ref{Sl}) we finally obtain the desired result
\begin{equation}
       \Delta^a S^{\{l+1\}}[A,\psi,\bar\psi] = 0.
\end{equation}
\hspace*{\fill} $\Box$

Of course, from eq. (\ref{ofcourse}) it also follows that
the weakly renormalized Action $S_w^{\{l\}}$ is gauge invariant.

\section{POWER COUNTING IN LIGHT CONE GAUGE} \label{4}
The main feature of Feynman graphs in light-cone gauge is the presence of
spurious poles introduced by the particular form of the free gauge field
propagator
\begin{eqnarray}
 \langle 0| T A^a_\mu (x) A^b_\nu (y) |0 \rangle_{g=0}
= \int {d^{2\omega}k\over (2\pi)^4} e^{ik(x-y)}
 \frac{-i\delta^{ab}}{k^2+i\epsilon} \left[
 g^{\mu\nu} - \frac{n^\mu k^\nu + n^\nu k^\mu}{[[nk]]} \right] \: .
\end{eqnarray}
The prescription of the spurious pole is the so called Mandelstam-Leibbrandt
(ML) prescription \cite{7,8,9}
\begin{equation}           \label{ML}
   {1\over [[nk]]} \stackrel{Man}{\equiv}
   \frac{1}{nk+i\epsilon \sigma(\ns k)}
   \stackrel{Lei}{\equiv}
   \frac{\ns k}{(nk)(\ns k)+ i \epsilon}\: ,
\end{equation}
being $\sigma(\cdot)$ the sign function and $\ns_\mu$ a new four-vector
on the light-cone independent from $n_\mu$. The choice of $\ns_\mu$
represents a further violation of Lorentz covariance.
We choose $n_\mu= {1\over \sqrt{2}} (1,0,0,1)$ and
$\ns_\mu= {1\over \sqrt{2}}(1,0,0,-1)$ in a particular Lorentz frame.
Therefore
\begin{equation}
    \frac{1}{[[nk]]} =  \frac{1}{k^+ + i \epsilon \sigma(k^-)}
\end{equation}
where we have introduced the light-cone coordinates (LCC)
\begin{equation}
	 k^\pm = k_\mp = \frac{k^0 \pm k^3}{\sqrt{2}} \: .
\end{equation}
One can prove that the derivative of the ML distribution with
respect to $p^\mu$ is
given by\cite{nota}
\begin{equation}      \label{der}
       \frac{\d}{\d p^\mu} \frac{1}{[[np]]} = - \frac{n_\mu}{[[np]]^2} -
2\pi i n^*_{\mu} \delta (np) \delta (n^*p) \: .
\end{equation}
This result will be useful when discussing the form of non-polynomial
counterterms in section \ref{5}.

The ML distribution has also correct homogeneity properties with respect
to both $n_\mu$ and $\ns_\mu$; this can be seen observing that
\begin{equation}      \label{omon}
     n^\mu \frac{\d}{\d n^\mu} \frac{1}{[[np]]} = - \frac{1}{[[np]]} \: ;
\end{equation}
while  eq. (\ref{ML}) is manifestly
invariant under dilation of the vector $\ns_\mu$.
As a consequence, the homogeneity degrees of a composite
operator with respect to both gauge vectors are preserved
under renormalization.

One can show\cite{6,17} that using the ML prescription, the euclidean UV
power counting is a good convergence criterion for the corresponding
minkowskian integrals. On the other hand, the spurious poles behave as
convergence factors only for the ``longitudinal'' variables $k^0$ and
$k^3$ and not for the ``transverse'' ones $k^1$ and $k^2$.
It follows that in light-cone gauge a diagram may have  a divergence
associated to certain proper subsets of the integration variables yet
involving all integration momenta. From an analytical point of view ---
i.e. as for Weinberg's theorem --- these divergences are
subdivergences, but from a graphical point of view they are to be
considered as overall divergences because they are not related to
subdiagrams and therefore they are not removed by counterterms
in the graph-by-graph   subtraction method.
Hence we shall call ``superficially divergent'' a graph $G$ if it
exhibits positive power counting on some subset (proper or not)
of its integration variables not limited to a proper subdiagram of $G$.
In the following we introduce an appropriate superficial degree of
divergence consistent with this definition.

First we consider a 1-loop diagram $G^{(1)}$. We denote with
$\delta_\forall (G^{(1)})$  the usual degree of divergence one obtains
by a dilation  of  all the variables $\{k^\forall\}= \{k^0,k^1,k^2,k^3\}$
and we define the analogous quantity $\delta_\perp (G^{(1)})$
obtained considering just the transverse variables
$\{k^\perp\}= \{k^1,k^2\}$.
$\delta_\perp$ differs from $\delta_\forall$ on differentials
\begin{equation}
	   \delta_\perp(d^4 k)= 2
\end{equation}
or in the cases
\begin{equation}
\delta_\perp\left( nk \right) =
\delta_\perp\left( \ns k \right) =
	   \delta_\perp\left(\frac{1}{[[nk]]}\right)= 0 \: ,
\end{equation}
while we shall keep for a single component
\begin{equation}
	   \delta_\perp(k^\mu)=\delta_\forall(k^\mu)= 1 \: ,
\end{equation}
as the result of integrals will always be written in
four-vector notation.

The ``superficial degree of divergence'' of $G^{(1)}$ is then defined as
\begin{equation}
	   \delta(G^{(1)}) = \max \left\{
	   \delta_\forall(G^{(1)}), \delta_\perp(G^{(1)}) \right\}.
\end{equation}
It is easy to show that $\delta(G^{(1)})$  is the maximum degree among
the ones related to all possible subsets of  integration variables.

Now we consider a graph $G$ with l integration  momenta $k_1, \ldots ,k_l$.
We still define
\begin{eqnarray}
       \{k_i^\forall\}&=& \{k_i^0,k_i^1,k_i^2,k_i^3\}   \\
       \{k_i^\perp\}&=& \{k_i^1,k_i^2,\}
 \hspace{1.5cm}
       i=1,\ldots,l, \: ,
\end{eqnarray}
and denote by
\begin{equation}
\delta_{v_1,\ldots,v_l}(G)  \hspace{2cm}  v_i \in \{ \forall,\perp\}  \: ,
\end{equation}
the degree of divergence of $G$ related to the variables
\begin{equation}
	   \{k_1^{v_1} \} \cup \{k_2^{v_2} \} \cup \cdots
	   \cup \{k_l^{v_l} \}  \: .
\end{equation}
The superficial degree of divergence of $G$ is now defined as
\begin{equation}
       \delta (G) = \max_{v_i \in \{ \forall,\perp\}}
       \left\{  \delta_{v_1,\ldots,v_l}(G) \right\} \: .
\end{equation}
It is easy to realize that this definition leads to a sufficient
condition for convergence. To show that it is not too cautious,
we look at the following two-loop example: for the integral
\widetext
\begin{equation}
  I = \int \!\! d^{2\omega}k_1\:d^{2\omega}k_2
  \frac{ k_1^\mu\, k_1^\nu\, k_2^\rho
  }{(k_1-q)^2 (k_1-k_2)^4  k_2^2
  \,[[n(k_1+k_2+s)]]
  \,[[n(k_1+p)]]
  \,[[nk_1]]^2},
\end{equation}
\narrowtext one finds
\begin{equation}
\delta (I)= \max
\left\{
   \begin{array}{llr}
     \delta_{\forall\forall} &=& -1 \\
     \delta_{\perp\forall} &=& 0 \\
     \delta_{\forall\perp} &=& -3 \\
     \delta_{\perp\perp} &=& -1
   \end{array} \right\}=0 ,
\end{equation}
and hence $I$ may diverge in the variables
$\{k_1^\perp \} \cup \{k_2^\forall \}$. We remark that $I$ has negative
mass dimension and no subdivergences.

We  say that a diagram $G$ is superficially convergent (divergent)
if $\delta(G)<0$ $(\delta(G)\geq 0)$. We say that $G$ has a
subdivergence if it has a superficially divergent 1PI proper subdiagram.

It is crucial to notice that, while in covariant gauges
the usual degree $\delta(G)$ has a dimensional meaning because it equals
the dimension of the 1PI momentum-space graph, in  light-cone gauge
$\delta(G)$  depends on the particular topological structure of the graph.
As a consequence, different graphs
contributing to the same proper vertex have in general a different degree
and any proper vertex, whatever its dimension, can have superficially
divergent graphs.
Therefore, in light-cone gauge power counting arguments do not limit
the type of counterterms entering the lagrangian or
a composite operator under renormalization. In particular, as even diagrams
with negative mass dimension may be superficially divergent,  non-local
--- i.e. non polynomial in external momenta --- counterterms are
generally expected.

\section{GENERAL FORM OF NON-POLYNOMIAL COUNTERTERMS} \label{5}

In a covariant field theory and in particular in Yang-Mills theories
with covariant gauges, the so called BPH theorem holds.
The counterterm $CG$ of a 1PI graph $G$ is polynomial in the external
momenta and thereby the locality of the lagrangian or a composite operator
is preserved under renormalization.
The theorem does not hold in light-cone gauge: the lacking argument
in the proof is that in a covariant theory, the action of the derivative with
respect to an external momentum on a graph $G$, lowers its degree of
divergence. In light-cone gauge this is not true. Consider for instance a
graph $G$ with $l$ integration momenta $k_1,\ldots,k_l$. Suppose that $G$
has a spurious pole of the form
\begin{equation}
   \frac{1}{[[n(k_1 + r + \mbox{other momenta})]]}
\end{equation}
r being an external momentum.
The degree $\delta(G)$ is not necessarily lowered by a differentiation
with respect to $r^\mu$, as
the degrees  $\delta_{\perp, v_2,\ldots,v_l}$ are not. As a  consequence,
$CG$ is not in general a polynomial in $r^\mu$.
However it is easy to see that a suitable number of  derivatives
\begin{equation}
	  \frac{\d}{\d p^\alpha}  \hspace{2cm}  \alpha \in \{-,1,2\}
\end{equation}
acting on a graph $G$, does indeed make it converge (see eq. (\ref{der})).
Therefore
the BPH theorem  is modified as follows:
\begin{prop} Let $G$ be a 1PI diagram in light-cone gauge and $p$ an external
momentum. If $\delta(G)\geq 0$, then $CG$ is a polynomial
in the components $p^\alpha$, $\alpha \in \{-,1,2\}$.
\end{prop}
Possible non-localities of counterterms are therefore limited to non
polynomial functions of $p^+_i=np_i$.
We shall see that the non localities can only
appear as spurious poles
$[[np]]^{-1}$ in the external momenta.

By the same arguments one can show that in light cone gauge,
as in  covariant gauges, for a 1PI graph $G$, the counterterm $CG$ is
polynomial in the fermionic masses.

Some results about the most general form of non-local counterterms
prove to be very important in renormalization theory.
The following proposition states that the only possible non-localities of a
counterterm are spurious poles $\frac{1}{np}$ in the external momenta.
In the proof, the following ``splitting formula'' holding for the ML
distribution, is used,
\begin{eqnarray}      \label{split}
\frac{1}{[[n(p_1+k)]]}\frac{1}{[[n(p_2+k)]]}=
\frac{1}{[[n(p_2-p_1)]]}
\left(
\frac{1}{[[n(p_1+k)]]} - \frac{1}{[[n(p_2+k)]]}
\right)  \: .
\end{eqnarray}
\begin{prop}         \label{prop1}
let $G$ be a 1PI graph in light-cone gauge. Without loss of generality
we can consider
\begin{equation}
     G = \int  \frac{d^{2\omega}k_1 \ldots d^{2\omega}k_l\:\:\:
         f(p,k,n,\ns,g_{\mu\nu})}{
         \prod_{j=1}^\alpha \left(t^2_j(k,p)-m^2_j \right)   \:
	 \prod_{k=1}^\beta [[ns_k(k,\ptilde)]]} ,
\end{equation}
where:
\begin{itemize}
\item $p_i$ are the external momenta;
\item $\ptilde_ j$ $(j=1, \ldots, \beta)$
are linear combination of the $p_i$;
\item $f$($\hat f$) is a polynomial in its arguments;
\item $t_j$($\hat t_j$) are linear combinations of the $p_i$ and $k_i$;
\item $s_k$($\hat s_k$) are linear combinations of the  $k_i$ and $\ptilde_i$.
\item $m_j$($\hat m_j$) are possible fermionic masses.
\end{itemize}
Then $G$ can be expressed as a sum
\begin{equation} \label{decomp}
    G= \sum_k  I_k,
\end{equation}
where each $I_k$ is of the form:
\begin{eqnarray}  \label{I}
I= \prod_{r=1}^\gamma \frac{1}{n\hat s_r(\ptilde)} \hat I =
   \prod_{r=1}^\gamma \frac{1}{n\hat s_r(\ptilde)}
   \int  \frac{d^{2\omega}k_1 \ldots d^{2\omega}k_l \:\:\:
         \hat f (p,k,n,\ns,g_{\mu\nu})}{
         \prod_{j=1}^\alpha \left(\hat t^{2}_j(k,p)
           -\hat m^{2}_j \right)   \:
         \prod_{m=1}^l  [[nk_m]]^{\beta_m}  },
\end{eqnarray}
with $\beta_m \geq 0\:\:\forall m$ and
\begin{equation}
    \sum_{m=1}^l \beta_m \: + \: \gamma= \beta\:\:\:\Rightarrow\:\:\:
    \gamma\leq\beta .
\end{equation}
\end{prop}
\begin{cor}     \label{cor1}
$CG$ is a meromorphic function in the variables
$p^+_i=np_i$ with poles at most of order $\beta$.
\end{cor}
{\bf proof:} (corollary)
it follows directly from the form of the integrals $\hat I$
observing that $C\hat I$ is polynomial in the external momenta
as the spurious poles are $p_i$-independent. \hspace*{\fill} $\Box$

\vspace{0.5cm}

\noindent{\bf proof:} (proposition) by induction on $l$.
We first show the thesis for $l=1$. If no spurious poles depend on
$k_1$, then $G$ is already of the form $I$ with $\beta_1=0$ and
$\gamma=\beta$.
Otherwise, by using formula (\ref{split}),
one can factor out of the integrand all spurious poles but one
 $[[n(\ptilde + k_1)]]^{-\beta_1}$ that can  possibly be of higher order
$(\beta_1>1)$ if multiple poles were originally present in $G$.
By shifting $k_1$ these poles become $[[nk_1]]^{-\beta_1}$ and
therefore $G$ is decomposed as in eq. (\ref{decomp}).

Let us now assume that the thesis holds for $l-1$ loops; we can apply
to the integral in  $d^{2\omega}k_l$ the same procedure above
considering as ``external momenta'' also the
variables $k_i$ $i=1, \ldots, l-1$.
\widetext $G$ is therefore reduced to a sum
of terms of the form
\begin{eqnarray}
\int\frac{   d^{2\omega}k_1 \ldots d^{2\omega}k_{l-1}
}{\prod_{r=1}^{\beta-\beta_l}
      [[ n\hat s_r(\ptilde,k_1,\ldots,k_{l-1})]]}
 \int
        \frac{d^{2\omega}k_l\: \hat f(p,k,n,g_{\mu\nu})}{
        \prod_{j=1}^\alpha
       \left(\hat t^{2}_j(k,p)-\hat m^{2}_j \right)   \:
           [[nk_l]]^{\beta_l}  } =  \nonumber
\end{eqnarray}
\begin{eqnarray}
 =\int \frac{d^{2\omega}k_l}{[[nk_l]]^{\beta_l}}
 \int
\frac{ d^{2\omega}k_1 \ldots d^{2\omega}k_{l-1} \:\:\:
\hat f(p,k,n,g_{\mu\nu})}{
         \prod_{j=1}^\alpha
 \left(\hat t^{2}_j(k,p)-\hat m^{2}_j \right)  \:\:
 \prod_{r=1}^{\beta-\beta_l}
[[n\hat s_r(\ptilde,k_1,\ldots,k_{l-1})]] }. \nonumber
\end{eqnarray}
\narrowtext
We can now apply the inductive hypothesis
to the multiple integral in the r.h.s.
considering $p_i$ and $k_l$ as external momenta but remembering
that the variables $\ptilde_i$ do not depend on $k_l$.
As a consequence, the spurious poles extracted from the multiple
integral can be factorized out of the integral in $d^{2\omega}k_l$
giving only terms of the form (\ref{I}). \hspace*{\fill} $\Box$

We now discuss a feature of Feynman integrals in light cone gauge
that is fundamental in selecting the possible structures
involved in operator renormalization. To this purpose let us consider
the following peculiar property of the ML prescription under the
algebraic splitting
\begin{eqnarray}
&&{1\over {nk+i\epsilon \sigma(n^*k)}}{1\over {n(k-s)+i\epsilon
\sigma(n^*(k-s))}}= \\
&& ={1\over {ns+i\epsilon \sigma(n^*s)}}\left [{1\over {n(k-s)+i\epsilon
\sigma(n^*(k-s))}}-
{1\over {nk+i\epsilon \sigma(n^*k)}}\right ]. \nonumber
\end{eqnarray}
In the limit $s_\mu \to 0$, no singularity is present in either term
of the equality; in particular at the left hand side we have a double
pole at $nk=0$ with causal prescription and at the right hand side the
pole at $ns=0$ is cancelled by a corresponding zero of the quantity in
square brackets. However, would we consider the limit $ns \to 0$ with
$n^*s \ne 0$, a singularity at $ns=0$ would persist owing to the
dependence on $n^*$.

This is at variance with the behaviour of a prescription involving only
one gauge vector, after the disposal of Poincar\'e-Bertrand terms
\cite{17},
and is at the root of the non-local behaviour of some counterterms in
light-cone gauge with the ML prescription.

It is however clear that, should we restrict the spurious denominators
to the subregions $n^*s_i=-ns_i$, $s_i$ being any generic four-vector,
we would get
\begin{equation}
{1\over {ns+i\epsilon \sigma(n^*s)}}
\to {ns\over (ns)^2-i\epsilon} = CPV {1\over
{ns}},
\end{equation}
and would recover locality by the very same argument which is used in the
proof in the one-vector space-like case \cite{17}. We stress that this
restriction must be understood in the sense of the theory of
distributions. In particular multiple poles should always be interpreted as
derivatives, otherwise one immediately runs into powers of CPV
prescription, i.e. meaningless quantities.

Having the above discussed property in mind, we now require that
all acceptable non-local structures in counterterms, have to become
local when $n^*\partial$ is replaced by $-n\partial$.

Actually a replacement of the kind $n^*\partial \rightarrow
\kappa \: n\partial$, with any constant $\kappa$, would do the job.
Our previous choice is reminiscent of the condition $n^*\rightarrow
-n$, advocated in ref. \cite{6} to recover locality in analogy with the
spacelike planar gauge. We stress however that the present condition
is imposed in a form of a phase-space restriction while standing
on the light-cone; in this sense it is closer to the spirit of the
discussion in ref. \cite{18}.

We shall show in the next Section that this criterion is extremely
efficient in selecting among {\it a priori} possible non-local
structures the only acceptable one: $ \Omega $, the same already
present when renormalizing the effective Action \cite{6}.

\section{CONSTRUCTION OF RENORMALIZATION CLASSES} \label{6}
In covariant Yang-Mills theories,
the renormalization of gauge invariant composite operators
is governed by the BPH theorem;
the locality of counterterms guarantees that
a composite operator can only
mix with operators of lower or equal canonical dimension. This
{\em dimensional hierarchy} automatically limits the number of
renormalization constants needed by a single renormalized operator.
Nevertheless, in covariant gauges, the renormalization of
composite operators is a very complicated matter because of the
presence of non physical degrees of freedom that contribute non trivially
to renormalized operators. For this reason gauge invariance
of a composite operator is generally lost under renormalization \cite{3}.

In light cone gauge the situation is opposite.
Renormalization preserves  gauge invariance of the operator, but
the  presence of non local counterterms could allow in principle
an infinite number of independent structures to appear.

Our aim is to show that on the contrary  the renormalization
of a gauge invariant composite operator involves only a finite number
of renormalization constants and that non local terms
do not affect physical quantities.
{}From now on we will focus on  weak renormalization (a single insertion)
because, in the more general case,
an infinite number of counterterms is expected on general grounds also
in covariant theories if the operator has dimension $\geq 4$.

Let us consider a local gauge invariant operator $X$
being a Lorentz tensor of rank $i$ ($i$ free Lorentz indices),
with homogeneity degrees $O_n$ and $O_\ns$ with respect to the gauge
vectors and mass dimension $d_m$.
The most general form of the renormalized operator $[X]$ is a structure
having the same characteristics of $X$ mentioned above but locality
since poles of the form $n\d^{-1}$ may be present; however
the structure has to become local if the substitution
$\ns\d \to n\d$ is performed. Such structures will be called
{\em quasi-local}.
We observe that the {\it canonical} dimension of the field $A_{\mu}$
cannot be defined in light-cone gauge as the UV behaviour of its
propagator does depend on the gauge vector; the only well-defined
dimension of operators in light-cone gauge is mass dimension.
As a consequence in the expression of the renormalized operator
$[X]$ we shall consider mass parameters as
part of mixed operators and shall work with
dimensionless renormalization constants.
Hence we can state the following:
\begin{prop}
Local or quasi-local gauge invariant composite operators with the same mass
dimension $d_m$, the same homogeneity degrees $O_n$ and $O_\ns$, and the
same tensorial rank $i$ form a class that is closed under renormalization.
\end{prop}
We want to show that each of these {\em renormalization classes}
contain a finite number of independent operators.
In the construction of quasi-local gauge
invariant structures, besides the usual covariant tensors, spinors
and derivatives, the following covariant non local integral operator
can be used as a building block:
\begin{equation}
   [nD^{-1}]^{ab} = \delta^{ab} \frac{1}{[[n\d]]} -
   g\, f^{acd} \frac{1}{[[n\d]]}
   \left\{ nA^c [nD^{-1}]^{db} \right\};
\end{equation}
the formula has to be understood recursively and can  easily be expanded
in powers of $g$.
Because of the negative mass dimension of $nD^{-1}$, for any
given composite operator, an infinite set of possible independent
counterterms can be obtained still satisfying the requirements of
correct mass dimension, homogeneity and tensorial structure.
On the contrary, only  very few structures containing $nD^{-1}$
become local when $\ns\d\to n\d$; by explicit construction
one realizes that the only acceptable ones are those in which
$nD^{-1}$ is carried by the following combination
trasforming in the adjoint representation:
\begin{equation}      \label{Omega1}
\Omega^a = \frac{n^\mu n^{*\nu}}{n^*n} [nD^{-1}]^{ab} F_{\mu\nu}^b .
\end{equation}
This structure is peculiar since $n_\mu \ns_\nu F^{\mu\nu}$
develops a factor $nD$ in the numerator when  $\ns\d\to n\d$:
\begin{eqnarray}
\left.\Omega\right|_{\ns\d\to n\d} &=& \frac{1}{n^*n} \:
 \frac{1}{nD} \: \left(n\d\, \ns A - n\d\,n A
-ig [nA,\ns A]\right) =  \\
&=&  \frac{1}{n^*n}\: \frac{1}{nD} \:
 \left(nD \,\ns A - nD\,n A \right)  =
 \frac{1}{n^*n} ( \ns A - n A ).
\end{eqnarray}
Such a condition is indeed stronger than the one considered in ref.
\cite{6}
as structures like
\begin{equation}
    n_\mu \ns_\nu F^{\mu\nu} \times (\mbox{non local}) ,
\end{equation}
or
\begin{equation}
\frac{1}{nD}\: \ns D \times (\mbox{anything}) ,
\end{equation}
become local if one replaces $\ns\to n$ but not if $\ns\d\to n\d$.

The crucial point here to observe is that $\Omega$
has positive mass dimension; hence, for any given operator
the number of possible independent
gauge invariant counterterms built from local covariant objects and
$\Omega$, is automatically limited by dimensionally arguments.
Equivalently, each renormalization class is finite.
Moreover, by  directly inspecting the expansion of $\Omega$,
\begin{eqnarray}     \label{Omega2}
 \Omega^a &=& \Omega_0^a + \sum_{k=1}^\infty g^k \Omega_k^a  , \\
 \Omega_0^a &=&  n^* A^a - \frac{n^*\d}{n\d} nA^a , \nonumber \\
 \Omega_k^a &=& (-1)^{k+1} f^{ab_k h_{k-1}} f^{h_{k-1}b_{k-1} h_{k-2}}
 \cdots  f^{h_2 b_2 h_1}  f^{h_1 b_1 c} \cdot   \nonumber \\
&\cdot&  \frac{1}{n\d} \left\{ nA^{b_{k}}
	 \frac{1}{n\d} \left\{ nA^{b_{k-1}}
	 \frac{1}{n\d} \left\{  \cdots
	 \frac{1}{n\d} \left\{ nA^{b_{1}}
	  \frac{n^*\d}{n\d} nA^c \right\} \cdots \right\}
	  \right. \right. ; \nonumber
\end{eqnarray}
one learns that all non local terms appearing in the renormalized operator
$[X]$ will be proportional to the field $nA$; therefore
only the local part of the
renormalized operator $[X]$ will contribute to $\lambda$-independent
Green's functions (see eq. (\ref{14})).

What we have said before can be summarized as follows:
\begin{prop}
Let $X$ be a local or quasi-local gauge invariant
composite operator; then
\begin{enumerate}
\item $[X]$ involves a finite number of renormalization constants
\item possible non local terms of $[X]$ are proportional to $nA$
and therefore do not contribute to physical quantities.
\end{enumerate}
\end{prop}

Let us see how to build a basis of independent operators for a given
renormalization class characterized by the mass dimension $d_m$, the
homogeneity degrees $O_n$ and $O_\ns$, and the tensor rank $i$ of
its gauge invariant operators. In the following table we list
the ``blocks'' that can be used to build a local or quasi-local
operator:
\begin{equation}  \label{68}
  \begin{array}{rcl}
     f &=& \# [ \bar\psi \cdots \psi ] \\
     q &=& \# [ F_{\mu\nu} = F_{\mu\nu}^a\tau^a ]  \\
     p &=& \#[D_\mu^{ab}   \:\:\:\: \mbox{\rm or}\:\:\:\:
		\D_\mu ]  \\
  \omega &=&\#[ \Omega=\Omega^a \tau^a ] \\
     j  &=& \#[n_\mu ]\\
     k &=& \#[n^*_\mu ]\\
     l &=& \#[ (n^*n)^{-1}] \\
     g &=& \#[\gamma^\mu ]\\
     r &=& \#[g^{\mu\nu}] \\
     m &=& \# [\mbox{\rm masses
           $m$ or derivatives $\d_\mu$ }]
   \end{array}
\end{equation}
where of course $g^{\mu\nu}$ is understood only with free indices.
The positive integer variables $f,q,\ldots,r$ denote the
multiplicity of a single factor inside a given operator.
Of course, as already anticipated  in (\ref{68}),
derivatives acting on gauge invariant quantities are also
allowed, each one entailing a unit dimension  in mass.
The values the variables can assume are subject to the  costraints due to
mass dimension and homogeneity respect to $n$ and $\ns$:
\begin{mathletters}
\begin{equation}                       \label{costra}
  3f + 2q + p + \omega + m  = d_m.
\end{equation}
\begin{equation}
  j-\omega-l = O_n,                    \label{costrb}
\end{equation}
\begin{equation}
  k - l = O_\ns,                       \label{costrc}
\end{equation}
\end{mathletters}
Eq. (\ref{costra}) gives an upper limit to all the variables in the l.h.s.;
the remaining variables are always limited by imposing the correct rank $i$
of the operators and their independence.

Of course more operators can correspond to the same combination
of variables. Starting from the table above, one has first to build
all possible combinations, whose number is however {\it finite} for
a given composite operator, and then to check their independence.
Some examples are discussed in the next Section.

\section{EXAMPLES} \label{7}
In this Section we discuss two simple examples of mixing.

The first gauge invariant composite operator we consider is
$\bar\psi \psi$. One can easily realize that the only
allowed counterterms are
\begin{equation}
[\bar\psi \psi]=\zeta_1 \bar\psi \psi+ \zeta_2 \bar\psi {n\!\!\!/
n^*\!\!\!\!\!/
\over{2nn^*}}\psi+\zeta_3 m^3.
\end{equation}
In this case renormalization
involves only local gauge invariant fermionic bilinears, besides
the mass term. An explicit one-loop calculation gives
\begin{eqnarray}
&&\zeta_1 = 1-{g^2 \over{8 \pi^2}}{N^2-1
\over{2N}}{1\over{2-\omega}}, \nonumber\\
&&\zeta_2 =0,\\
&&\zeta_3={N\over{4\pi^2}}{1\over{2-\omega}}. \nonumber
\end{eqnarray}

As a second example we consider the fermionic U(1) conserved current
$\bar\psi \gamma_{\mu} \psi$. There are several independent
gauge invariant structures with mass dimension 3 it can {\it a priori}
mix with:
\begin{eqnarray}
[\bar\psi \gamma_{\mu} \psi]=\zeta_1 \, \bar\psi \gamma_{\mu} \psi +
\zeta_2\,{n_{\mu}\over {nn^*}}\, \bar\psi n^*\!\!\!\!\!/ \;\psi +
\zeta_3\,{n^*_{\mu}\over {nn^*}}\,\bar\psi n\!\!\!/ \psi + \\
+\zeta_4 \, \bar\psi{n\!\!\!/  \gamma_{\mu} n^*\!\!\!\!\!/ \over {nn^*}}
\psi + \zeta_5\, n^{\nu}F_{\nu \mu}^a\Omega^a + \zeta_6\, n_{\mu}
{n_{\nu}n^*_{\sigma}\over {nn^*}}F^{\nu \sigma,a}\Omega^a. \nonumber
\end{eqnarray}
Some of them are non local, i.e. involve $\Omega$. However,
this current is related to the fermion propagator by the U(1) Ward identity
which sets constraints between $\zeta_i$ 's and the wave function
renormalization constants $Z_2$ and $\tilde {Z}_2$ \cite{6}.
An explicit one-loop calculation
fully confirms this result and reproduces the values \cite{6}
\begin{eqnarray}
&&\zeta_1=Z_2^{-1}=1-{g^2 \over{16 \pi^2}}{N^2-1
\over{2N}}{1\over{2-\omega}}, \nonumber \\
&&\zeta_2=-\zeta_3=\tilde {Z}_2^{-1}-1 =
{g^2 \over{8 \pi^2}}{N^2-1 \over{2N}}{1\over{2-\omega}},\\
&& \zeta_4 = \zeta_5 = \zeta_6 = 0. \nonumber
\end{eqnarray}

Finally a last example, definitely requiring non-local counterterms, is
the lagrangian density itself, as discussed in \cite{6}; of course
further structures involving total derivatives must be considered
in this case.

\section{CONCLUSIONS} \label{8}
In this paper we have solved the problem of renormalizing at any
order in the loop expansion gauge invariant composite operators in
Yang-Mills theories quantized in light-cone gauge with the correct
causal ML prescription in vector propagator.
We have here generalized the treatment developed in refs.\cite{6,17},
concerning effective Action.

Main results are:
\begin{itemize}
\item the proof that renormalization preserves the gauge invariance of
composite operators (Sect. \ref{3}).
\item the full characterization of admissible non-local structures
in counterterms, which can be only carried by the
quantity $\Omega$, and therefore cannot contribute to physical
quantities; hence the proof that the renormalization of a composite
operator always entails a finite number of renormalization constants
(Sect. \ref{6}).
\end{itemize}
Specific examples of mixing under renormalization
are presented in Sect. \ref{7}; in particular we have found
quite instructive the behaviour of the U(1) conserved
fermionic current, which is endowed with a direct physical interest.

Generalizations to gauge dependent operators are {\it a priori}
possible,
however one is immediately faced with a basic difficulty concerning
Lee-Ward identities, which are no longer form-invariant under
renormalization. Besides, physical applications are usually concerned
with gauge invariant composite operators.

Few preliminary results have already appeared
in the literature  \cite{19}, while a pedagogical review on the
whole subject will be reported elsewhere  \cite{20}.


\begin{references}
\bibitem[*]{nota} We take here the opportunity of correcting eq. (A6.2) of
reference \cite{17}.
\end{references}
\end{document}